\documentclass[letter]{aa}   
\usepackage{CJKutf8}
\usepackage{graphicx}
\usepackage[breaklinks,colorlinks,citecolor=blue,linkcolor=blue]{hyperref}
\usepackage{txfonts}
\usepackage{lscape}
\usepackage{epstopdf}
\usepackage{CJK}
\usepackage{float}
\usepackage[absolute]{textpos}
\usepackage{amsmath}
\usepackage{mathrsfs}

\usepackage{arydshln}

\begin{document} 

   \title{Kolmogorov turbulence across multi-fractal gas in Polaris Flare}


%
%
%

   \author{Xunchuan Liu$^1$\thanks{liuxunchuan001@gmail.com} \and 
    Pak-Shing Li$^2$ \and Yihuan Di$^{3,2}$
        }

   \institute{Leiden Observatory, Leiden University, P.O. Box 9513, 2300RA Leiden, The Netherlands 
   \and 
   Shanghai Astronomical Observatory, Chinese Academy of Sciences, Shanghai 200030, PR China
   \and 
   School of Physics and Astronomy, Shanghai Jiao Tong University, Shanghai 200240, PR China
   }

\abstract{We reveal a pristine, scale-invariant 3D Kolmogorov velocity cascade 
($\alpha_V^{\mathrm{3D}} \sim 2/3$) spanning $0.05$--$20$~pc in the Polaris 
Flare using \texttt{PPCOS} $^{12}\text{CO}$ data. A transition scale at 
$\sim 0.5$~pc marks a bifurcation in the structure functions' exponents,
below which the degree of intermittency is also saturated. By deriving an 
analytical mapping relation ($\alpha_V^{\mathrm{3D}} = \alpha_V - 
\frac{1}{3}\alpha_I$), we obtain the scale-invariant value of 
$\alpha_V^{\mathrm{3D}}$, proving that the apparent transition stems from 
geometric projection and a changing density fractal dimension rather than 
a turbulent mode shift. Kolmogorov turbulence is smoothly inherited from 
the large-scale cold neutral medium, remaining uninterrupted by compression or gravity
below 0.1 pc.
}



   \maketitle
    \nolinenumbers

\section{Introduction} \label{sec:intro}

Interstellar turbulence fundamentally regulates the lifecycle of gas clouds 
by providing macroscopic structural support, driving multi-scale energy dissipation, 
and compressing gas into highly localized structures 
\citep[e.g.,][]{Larson81,2004ARA&A..42..211E,2010A&A...512A..81F, 2025NatAs...9.1366L, 2025arXiv250220458L}. 
These dynamics serve as a critical gateway for understanding general cloud physics 
and large-scale Galactic architectures \citep[e.g.,][]{Spitzer1978, 1998ApJ...498..541K, 
2009ApJ...693..216K, 2016A&A...595A..37K}. Characterizing this turbulence 
requires measuring the spatial scaling behavior of the gas fields via the power-law 
exponent ($\alpha$) of the second-order structure function, 
$S_2(L) = \langle |f(\vec{x} + \vec{L}) - f(\vec{x})|^2 \rangle \propto L^\alpha$ 
\citep[e.g.,][]{1941DoSSR..30..301K, 2004ApJ...615L..45H}. Discerning the true 
three-dimensional exponent $\alpha_{V}^{\text{3D}}$ for the velocity field ($V$) is crucial 
to identifying the dominant turbulent pattern, separating classic incompressible 
cascades ($\alpha_{V}^{\text{3D}} = 2/3$; \citealt{1941DoSSR..30..301K}, hereafter K41) 
from highly compressible, shock-dominated environments ($\alpha_{V}^{\text{3D}} = 1$; 
\citealt{BURGERS1948171}), as well as smooth laminar flows governed by stable 
velocity gradients ($\alpha_{V}^{\text{3D}} = 2$). However, recovering these 
intrinsic three-dimensional scaling properties from astronomical observations is 
historically constrained by line-of-sight projection effects and a limited 
spatial map size relative to the achievable resolution 
\citep[e.g.,][]{2008A&A...485..917O, 2011A&A...529A...1S}.

The Polaris Flare, located at a high Galactic latitude ($b>20^\circ$) with a distance of $\sim$150~pc \citep{1998A&A...331..669F}, provides an ideal cosmic laboratory 
for studying interstellar turbulence \citep[e.g.,][]{2008A&A...485..917O, 2010A&A...518L.104M}. 
The PMO Polaris CO survey (\texttt{PPCOS}; \citealt{2026arXiv260618637L}, hereafter Paper~I) delivers a 100~$\text{deg}^2$ 
footprint ($\sim$10$^\circ$ in size, corresponding to a physical scale of $\sim$26~pc) fully Nyquist-sampled at sub-arcminute ($50\arcsec$) resolution towards the Polaris Flare, 
capturing large-scale energy injection at cloud boundaries while resolving the inertial cascade down to $\sim$0.04~pc (approximately 10\,000~au).
The diffuse outskirts of the Polaris Flare display a highly coherent dynamical coupling with the surrounding 
Cold Neutral Medium (CNM) \citep[][hereafter Paper~II]{2026arXiv260619694L}. 
Crucially, the cloud remains entirely untouched by active star formation \citep{2010A&A...518L.102A}, carrying cold dust temperatures, optically thin 
gas envelopes, and a simple energy injection profile \citep[e.g.,][see also Paper~I]{1998A&A...333..709L, 1999A&A...347..640B}.
Thus, \texttt{PPCOS} provides an unprecedented window into these turbulence-related physics.

In this letter, we apply $\Delta$-variance ($\sigma^2_\Delta$) analysis to the $^{12}$CO ($J=1-0$) integrated intensity map ($I$) and centroid velocity map ($V$) of the Polaris Flare. The measured spectrum of $\sigma^2_\Delta$ serves as a robust proxy for the second-order structure function slope $\alpha$ when $\alpha>0$ \citep{1998A&A...336..697S, 2002A&A...390..307O}. By co-analyzing the two-dimensional indices $\alpha_V^{\text{2D}}$ and $\alpha_I^{\text{2D}}$, we break the degeneracy between projection effects and spatial geometry to recover the true three-dimensional velocity scaling $\alpha_v^{\text{3D}}$ (with the method detailed in Sect.~\ref{sec_method}). Remarkably, this dual-index framework uncovers a classic Kolmogorov scaling ($\alpha_V^{\text{3D}} \sim 2/3$) extending across a vast spatial dynamic range enabled by \texttt{PPCOS}. We report this finding and its physical implications in Sect.~\ref{sec_turbpol}, validate it by analyzing the high-order features of the velocity structure in Sect.~\ref{sec_highorder}, and present our discussion in Sect.~\ref{sec_dis}. A breif summary is presented in Sect. \ref{sec_sum}.

\begin{figure}
    \centering
    \includegraphics[width=0.96\linewidth]{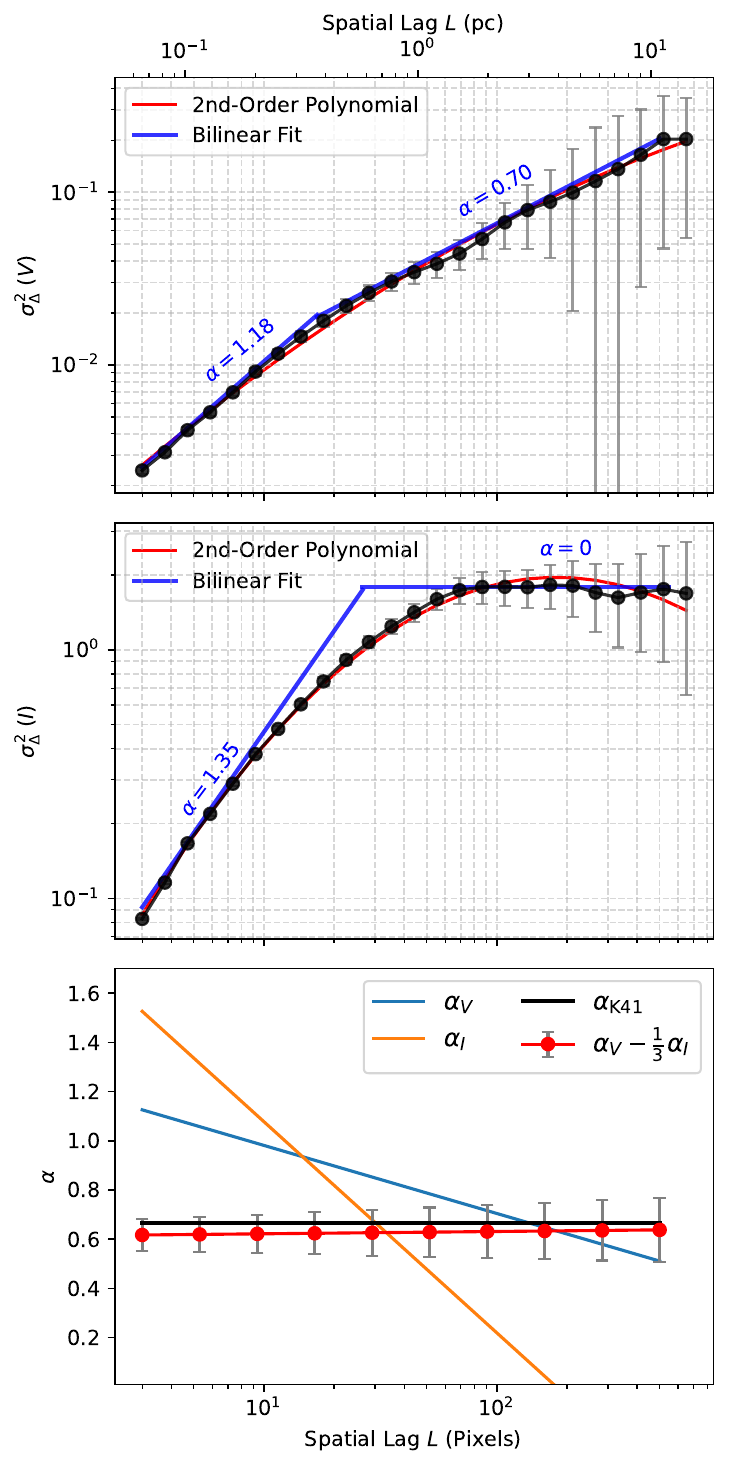}
\caption{Upper and Middle: $\Delta$-variance ($\sigma^2_\Delta$) spectra of the $^{12}\mathrm{CO}$ velocity ($V$) and integrated intensity ($I$) maps of the Polaris Flare (black dots) with bilinear power-law fits (blue lines) and second-order polynomial fits (red lines). The $\alpha$ values obtained from the bilinear fitting are indicated in blue (Sect. \ref{sec_turbpol}). Lower: Scale-dependent exponents ($\alpha$) derived from the first derivative of the polynomial fits. The red curve shows the inverted 3D velocity exponent ($\alpha_V^{\mathrm{3D}} = \alpha_V - \frac{1}{3}\alpha_I$), collapsing onto the K41 value $\alpha = 2/3$ (black).}
    \label{fig_alpha}
\end{figure}

\section{Method} \label{sec_method}

The \texttt{TurbuStat} package \citep{Koch2019AJ....158....1K} implements the $\Delta$-variance framework following \citet{1998A&A...336..697S} and \citet{2008A&A...485..917O}. A two-dimensional map is convolved with an azimuthally symmetric, zero-mean Mexican-hat wavelet kernel $\psi_L$ across spatial lag scales $L$, with the spectrum $\sigma^2_\Delta(L)$ computed as the mean square of the transformed map. This method is robust against irregular boundaries and observational noise (through smoothing).

The classic line-of-sight projection relation,
\begin{equation}
    \alpha^{\mathrm{2D}} = \alpha^{\mathrm{3D}} + 1, \label{eq_original}
\end{equation}
assumes a continuous, space-filling medium where integration along the line of 
sight increases the scaling exponent by exactly one dimension \citep{1981grf..book.....A,1998A&A...336..697S}. 
Real molecular clouds, however, are highly intermittent and possess a fractal 
topology, acting as low-dimensional manifolds ($D_c < 3$, and typically $D_c \gtrsim 2$) 
embedded in 3D space. 

A one-dimensional stochastic process with Hurst parameter $H$ exhibits a 
$\Delta$-variance exponent of $\alpha = 2H$, a relation that holds generally 
without requiring Gaussianity or strict stationarity. The Hausdorff dimension of its 
zero-crossing set is $D_{\mathrm{cross}} = 1 - H$, which implies that the fractal 
dimension of its active spatial support is $D_{\mathrm{active}} = 1 - 
D_{\mathrm{cross}} = H$ \citep{mandelbrot1968,RAMBALDI199421,biagini2008}. This active 
support conceptually represents the structural ``skeleton'' of the field. For a 
cloud embedded in 3D, line-of-sight integration restricts the scaling behavior according 
to this sparse skeleton (with a dimension of $D^{\rm skel}=2+H_\rho^{\mathrm{skel}}$) 
rather than a continuous volume, yielding the projection scaling rule 
$\alpha_\Sigma^{\mathrm{2D}} = \alpha_\rho^{\mathrm{skel}} + H_\rho^{\mathrm{skel}}$. 
Since $H_\rho^{\mathrm{skel}} = \alpha_\rho^{\mathrm{skel}}/2$ for a stochastic process, 
the 2D projection yields $\alpha_\Sigma^{\mathrm{2D}} = \alpha_\rho^{\mathrm{skel}} + 
H_\rho^{\mathrm{skel}} = \frac{3}{2}\alpha_\rho^{\mathrm{skel}}$, leading directly to the 
solutions\footnote{Notably, this formulation is formally self-consistent. Embedding the cloud skeleton into the 3D space (a reverse projection) leads to $\alpha_\rho^{\mathrm{3D}} = \alpha_\rho^{\rm skel} - (3 - D^{\rm skel}) = \frac{3}{2}\alpha_\rho^{\rm skel} - 1$. Applying the classic space-filling projection to return to 2D then yields $\alpha_\Sigma^{\mathrm{2D}} = \alpha_\rho^{\mathrm{3D}} + 1 = \frac{3}{2}\alpha_\rho^{\rm skel}$.} $\alpha_\rho^{\mathrm{skel}} = \frac{2}{3}\alpha_\Sigma^{\mathrm{2D}}$ and 
$H_\rho^{\mathrm{skel}} = \frac{1}{3}\alpha_\Sigma^{\mathrm{2D}}$. Here, we denote density as $\rho$ and column density as $\Sigma$.

Extending this framework to kinematics, the intensity-weighting of the centroid 
velocity (first-moment) map tightly couples the $V$ and $\rho$ fields. The mapping 
relation for the 3D velocity exponent, applicable both to the skeleton itself 
($\alpha^{\mathrm{skel}}_V$) and its embedding 3D space ($\alpha^{\mathrm{3D}}_V$), 
becomes:
\begin{equation}
    \alpha^{\mathrm{3D}}_{V} = \alpha^{\mathrm{2D}}_{V} - \frac{1}{3}\alpha_\Sigma^{\mathrm{2D}}. \label{eq_revi}
\end{equation}
Note that, because the physical velocity field persists continuously throughout the embedding 3D space 
even in regions devoid of the emission tracer, there is no dimensional reduction 
($\alpha^{\mathrm{3D}}_V = \alpha^{\mathrm{skel}}_V$). The relation
of Eq. \ref{eq_revi} is also supported by 
numerical simulations in Appendix~\ref{sec_simups}. Hereafter, the ``2D'' superscript 
is omitted where no ambiguity arises.

\section{Transition of turbulence in Polaris Flare}\label{sec_turbpol}
We apply the standard $\Delta$-variance fitting framework (Sect. \ref{sec_method}) to the $^{12}\mathrm{CO}$~(1--0) integrated intensity ($I$, as a tracer of  $\Sigma$) and centroid velocity ($V$) maps of the Polaris Flare delivered by the \texttt{PPCOS} survey (with a pixel size of 0.5\arcmin, Paper I). It is worth noting that the physical component traced by $^{12}\mathrm{CO}$~(1--0) 
corresponds to the $^{12}\mathrm{CO}$-excited molecular gas phase. 

\subsection{Intrinsic transition of velocity field?} \label{sec_clas}

As shown in Figure~\ref{fig_alpha}, both $\sigma^2_\Delta(I)$ and $\sigma^2_\Delta(V)$ display power-law behaviors as a function of the spatial lag ($L$) on log-log scales. Both spectra exhibit a steeper linear growth at small spatial scales ($L < 0.5\ \text{pc}$), which transitions smoothly into a shallower, flatter linear growth regime at larger spatial scales ($L > 0.5\ \text{pc}$).

A broken power-law fit to $\sigma^2_\Delta(V)$ yields scaling exponents of 
$\alpha_V = 1.18 \pm 0.03$ in the small-scale regime and $\alpha_V = 0.70 \pm 0.02$ 
in the large-scale regime. For $\sigma^2_\Delta(I)$, a linear fit at low $L$ 
yields $\alpha_I = 1.35\pm 0.06$, whereas $\alpha_I = 0$ provides a solid approximation 
at high $L$. Notably, the transition scale ($L_{\text{trans}}$) of $0.5~\text{pc}$ 
($\sim 11.5^\prime$) is comparable to the spatial resolution of the CfA CO survey. 
Based on these data, \citet{1998A&A...336..697S} reported a single $\alpha_V$ 
value of $0.8$, which closely aligns with our measured large-scale exponent. 
This agreement implies that the transition is physical—whether reflecting a change 
in the intrinsic Hurst mode or a geometric projection effect—rather than an 
artifact of limited image size.

If this transition in $\alpha_V$ is driven by a change in the intrinsic 
Hurst exponent, it points to a scenario where the system transitions from standard 
Kolmogorov turbulence at scales larger than $L_{\text{trans}}$, where self-gravity 
is negligible, to shock-dominated Burgers turbulence at smaller scales. 
 Notably, $L_{\rm trans}$ closely matches the spatial width ($0.25$--$0.5~\text{pc}$) 
of MCLD~123.5+24.9, the densest structure within the cloud complex 
\citep{2016MNRAS.462.1517P}.
This cascade shift below $L_{\text{trans}}$ likely occurs as self-gravity begins to play a role in structuring the dense gas, or where dissipation 
induced by shock compression takes effects.

\subsection{Transition of cloud fractal dimension?}\label{eq_reint}

However, the coupling of the spectral emission line with the highly non-uniform, fractal density field and line-of-sight projection effects (Sect.~\ref{sec_method}) fundamentally reshapes this classical interpretation. Instead, the transition likely reflects a scale-dependent shift in the fractal dimension of the underlying 3D skeleton of the fields. To capture this continuous transition, we apply a global second-order polynomial fit to the $\Delta$-variance spectra on log-log scales, yielding:
\begin{align}
   & \log_{10}\left[\sigma_\Delta^2(V)\right] = -0.14 x^2 + 1.25 x + C_V, &\label{eqV}\\
   & \log_{10}\left[\sigma_\Delta^2(I)\right] = -0.43 x^2 + 1.94 x + C_I, & \label{eqI}
\end{align}
where $x = \log_{10}(L/\text{pixel})$ (see upper and middle panels of Figure~\ref{fig_alpha}). The scale-dependent scaling exponents are subsequently recovered by taking the first derivative with respect to $x$, such that $\alpha(x) = d\log_{10}(\sigma_\Delta^2)/dx$:
\begin{align}
    \alpha_V(x) &= 1.25 - 0.28x, \label{aeqV}\\
    \alpha_I(x) &= 1.94 - 0.86x. \label{aeqI}
\end{align}

Notably, the derivative coefficient of $-0.28$ in Eq.~(\ref{aeqV}) is approximately 
one-third of the corresponding coefficient ($-0.86$) in Eq.~(\ref{aeqI}). When 
applying Eq.~(\ref{eq_revi}) to extract the intrinsic 3D velocity scaling exponent, 
these scale-dependent linear terms cancel each other out. This mutual cancellation 
effectively removes the scale dependence, yielding a nearly scale-invariant 
value of $\alpha_V^{\mathrm{3D}} \approx 0.62$--$0.64$ across the spatial 
scales traced by \texttt{PPCOS} ($0.05$--$20~\text{pc}$; Figure~\ref{fig_alpha}). 
This invariant behavior likely extends to both larger and smaller scales, and 
the recovered exponent is remarkably close to the classical Kolmogorov value of $2/3$.

\begin{figure}[t]
    \centering
    \includegraphics[width=0.98\linewidth]{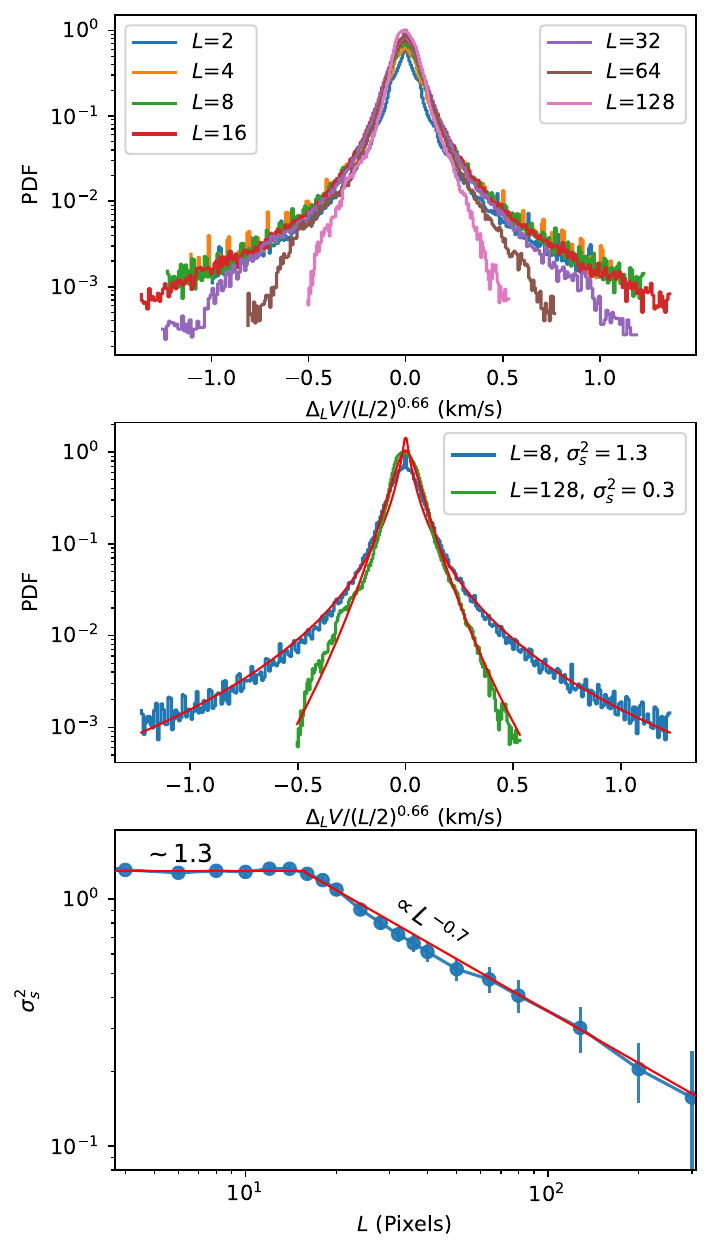}
    \caption{Top panel: PDFs of velocity increments $\Delta_L V$ at various lag scales $L$. The values of $\Delta_L V$ have been normalized by $(L/2)^{0.66}$ to compensate for the systematic broadening across different lag scales (Sect.~\ref{sec_highorder}). Middle panel: NLN fits (red curves) to two representative $\Delta_L V$ PDFs, with the best-fit log-variance $\sigma_s^2$ values indicated in the legend (Sect.~\ref{sec_highorder}). Bottom panel: Log-variance $\sigma_s^2$ as a function of lag scale $L$ (blue line) along with the corresponding power-law fit (red line).}
    \label{fig:skew}
\end{figure}

We suggest that the entire cloud system is embedded within a broader turbulent 
field traced by CO-dark gas and the cold neutral medium (CNM; Paper II) at scales 
exceeding $20~\text{pc}$, from which the internal turbulent cascade is inherited. 
At scales above $L_{\text{trans}}$, the cloud exhibits a fractal dimension of 
approximately $2$, which corresponds to a classical power-law density profile 
of $\rho \propto R^{-1}$ \citep{Larson81}. However, the underlying turbulent 
velocity pattern is smoothly transferred down to these smaller scales. 
Consequently, the intrinsic, volume-weighted velocity structure function remains 
completely unaffected by this localized structural mass reorganization.

\section{Fractal transition  characterized by intermittency}\label{sec_highorder}
The previous section focused on the second-order structure function. For a Gaussian 
process, the scaling exponent of the $n\text{-th}$-order structure function follows 
$\alpha_{n}=nH$. However, this linear relationship breaks down in turbulent flows 
due to intermittency driven by localized fluctuations in energy dissipation. As a 
result, the probability density function (PDF) of the two-point velocity increment, 
$\Delta_L V$, deviates from a Gaussian profile and exhibits heavy tails 
(Figure~\ref{fig:skew}). Consequently, the second-order moment no longer provides 
a complete statistical description. Structurally, the width of the central core 
of the $\Delta_L V$ PDF is proportional to $L^{0.66}$, which directly corresponds 
to the scaling exponent of the first-order structure function.

The distribution of $\Delta_L V$ is modeled as a continuous Normal–Log-Normal 
variance mixture (hereafter referred to as the NLN function). This approach is 
physically rooted in the refined similarity hypothesis of turbulence 
\citep[K62;][]{1962JFM....13...82K,1990PhyD...46..177C}, which posits that 
small-scale intermittency introduces log-normal fluctuations into the local energy 
dissipation rate. Specifically, we treat the local PDF of $\Delta_L V$ as a 
conditional Gaussian distribution:
\begin{equation}
    P(\Delta_L V \mid \sigma_{\Delta_L V}) = \frac{1}{\sqrt{2\pi}\sigma_{\Delta_L V}} \exp\left( -\frac{(\Delta_L V)^2}{2\sigma_{\Delta_L V}^2} \right),
\end{equation}
where the fluctuating standard deviation, $\sigma_{\Delta_L V}$, acts as a proxy 
for the spatial modulation of energy and follows a log-normal distribution \citep[as in density field, e.g.,][]{1994ApJ...423..681V,1997MNRAS.288..145P,2025arXiv250220458L}:
\begin{equation}
    P(\sigma_{\Delta_L V}) = \frac{1}{\sqrt{2\pi}\sigma_s \sigma_{\Delta_L V}} \exp\left( -\frac{(\ln\sigma_{\Delta_L V} - \mu)^2}{2\sigma_s^2} \right).
\end{equation}
Here, $\mu$ and $\sigma_s$ represent the mean and standard deviation of 
$\ln(\sigma_{\Delta_L V})$, respectively.

The NLN function provides a robust fit to both the sharp central peak and the heavy tails of the $\Delta_L V$ distribution (Figure~\ref{fig:skew}). 
Above the transition scale $L_{\text{trans}}$, the log-variance scales as $\sigma_{s}^2 \propto L^{-0.7}$, whereas it saturates at $\sim 1.3$ below this threshold. 
In contrast, the central core of the $\Delta_L V$ PDF maintains its $\propto L^{0.66}$ scaling across both regimes. 
This behavior points to two possible physical scenarios. 
First, the 3D intermittency pattern may remain strictly scale-invariant but appear saturated at small scales due to line-of-sight projection effects associated with the cloud's modified fractal dimension. 
Alternatively, this trend may reflect a genuine dynamical transition where the intrinsic intermittency itself is physically altered. 
Crucially, regardless of the underlying mechanism, the low-order (first- and second-order) structure functions remain invariant across all scales.

Above the $L_{\text{trans}}$ threshold, the power-law decay of $\sigma_{s}^2$ maintains 
a considerable value (e.g., $\sim 0.15$ at $L > 300~\text{pixels}$, corresponding to 
$\sim 6.5~\text{pc}$, the characteristic size of the $^{12}\text{CO}$ emission region) 
without any indication of a sudden drop. This sustained log-normal variance implies that 
turbulent energy is injected at or above these larger scales, further supporting 
our interpretation (Sect.~\ref{eq_reint}) that the turbulence within the Polaris 
Flare is inherited from the surrounding large-scale CNM.


\section{Limitations and discussions}\label{sec_dis}
Although our framework does not strictly require the $^{12}\mathrm{CO}$ emission to maintain a constant excitation temperature, we assume that the line is optically thin to preserve geometric symmetry between the line-of-sight and plane-of-sky dimensions. The highly diffuse nature of the Polaris Flare satisfies this criterion exceptionally well compared to more evolved, star-forming molecular clouds. However, this optically thin assumption may break down within the densest structures of the cloud core (see Paper~I), potentially contributing to the observed scaling turnover in $\sigma^2_\Delta$ around $L_{\rm trans}$.

Another underlying assumption is that $D_c \gtrsim 2$, which is physically 
justified at our spatial resolution and consistent with the lack of active star 
formation in the Polaris Flare (Paper~II). For sparse systems with $D_c < 2$, 
projection effects are statistically minimal due to negligible line-of-sight overlap. 
While our $0.05\ \text{pc}$ resolution resolves typical filament widths, higher-resolution 
observations are needed to trace the lower boundary of the K41 cascade. 
Limitations remain: first, the projection mapping for the log-variance $\sigma_s^2$ 
is currently unquantified, restricting high-order analysis to a qualitative assessment. 
Second, our model omits the interstellar magnetic field. Strong magnetic fields break isotropic symmetry via anisotropic turbulent channeling and induce Alfvénic coherence, causing the velocity field to deviate locally from an isotropic stochastic process.

\section{Summary}\label{sec_sum}
Using wide-field $^{12}\text{CO}$ ($J=1-0$) data from the \texttt{PPCOS} survey, 
we show that the Polaris Flare exhibits a remarkably scale-invariant, 
intrinsic 3D velocity scaling ($\alpha_V^{\mathrm{3D}} \sim 0.62$--$0.64$) 
that matches Kolmogorov turbulence across more than two orders of 
magnitude ($0.05$--$20~\text{pc}$). While localized mass restructuring alters the 
cloud's fractal geometry and saturates the intermittency log-variance ($\sigma_s^2 \sim 1.3$) 
below $0.5~\text{pc}$, the fundamental kinetic cascade remains uninterrupted. 
We conclude that the internal turbulence of the Polaris Flare is cleanly inherited 
from the surrounding large-scale CNM, maintaining a stable, scale-invariant 
energy cascade despite localized mass restructuring.

\begin{acknowledgement}
X.L. acknowledges the support of the Strategic Priority Research Program of the Chinese Academy of Sciences  under Grant No. XDB0800303.
We thank the staff of the Delingha Observatory for carrying out the observations. We thank Prof. Paul F. Goldsmith for helpful comments.
\end{acknowledgement}

\bibliography{PMOPolarisTurb}
\bibliographystyle{aa}

\clearpage
\begin{appendix}
\nolinenumbers

\section{Numerical simulation tests}\label{sec_simups}
We propose the quantitative relation 
$\alpha_V^{\mathrm{3D}} = \alpha_V - \frac{1}{3}\alpha_I$ 
(Eq.~\ref{eq_revi}), under the assumption that the density field does not 
deviate significantly from a random process. This formulation additionally 
depends on the underlying correlation between the velocity and density fields. 
Consequently, Eq.~\ref{eq_revi} is expected to provide a reliable 
estimation primarily for unevolved gas; it may break down in active 
star-forming regions where density field singularities emerge due to 
gravitational collapse. Furthermore, localized stellar outflow feedback 
may drive a complex turbulent 
state.
\begin{figure}[h]
    \centering
    \includegraphics[width=0.91\linewidth]{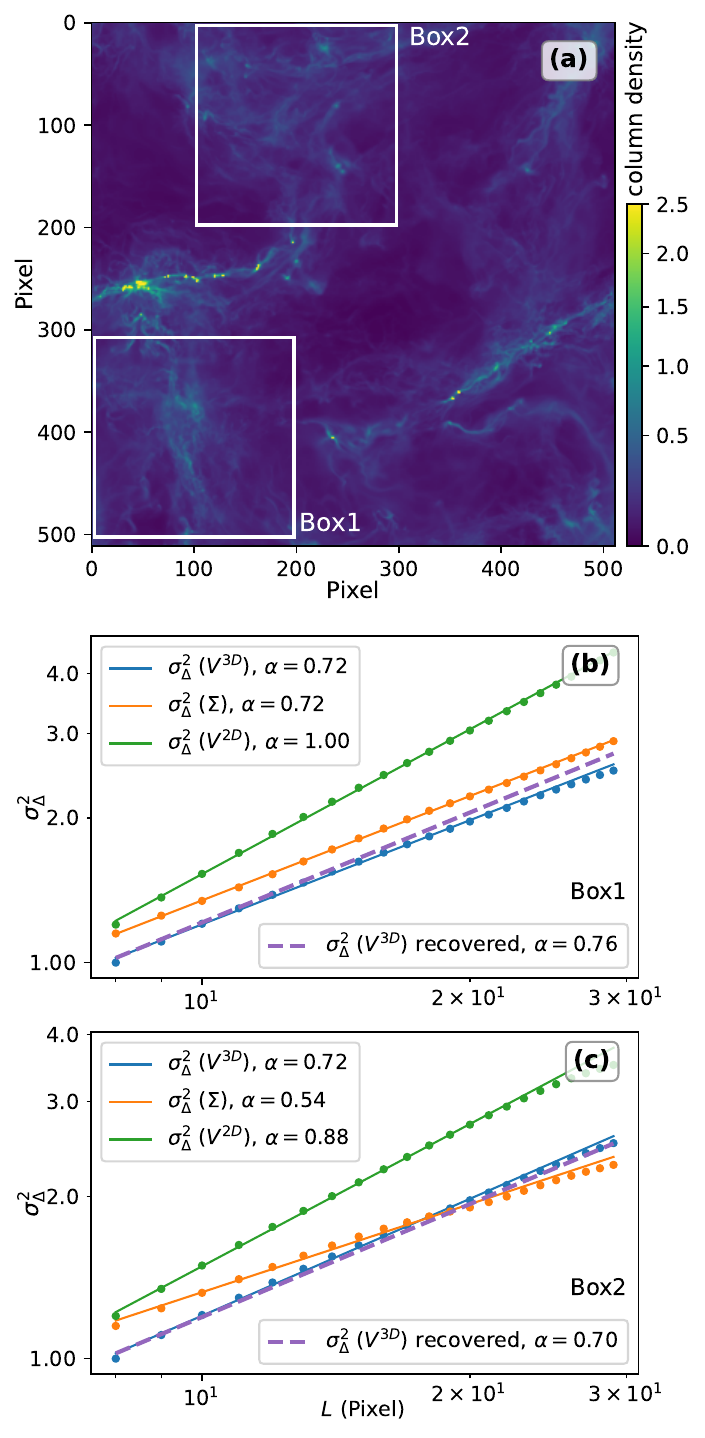}
    \caption{Panel (a): Column density map from numerical simulation conducted by \citet{2019MNRAS.485.4509L}. Two subregions unaffected by strong star-formation activity are marked by white boxes (Sect. \ref{sec_arx_si}). Panels (b--c): Comparison of the $\sigma^2_\Delta$ calculated directly from the 3D velocity cube versus that derived from the corresponding $V$ and $\Sigma$ maps (Appendix~\ref{sec_simups}) within the two boxes in panel (a) using Eq.~\ref{eq_revi}. 
    The solid lines in panels (b) and (c) represent linear fits in log--log space.}
    \label{fig:simups}
\end{figure}    

\begin{figure}[t]
    \centering
    \includegraphics[width=0.9\linewidth]{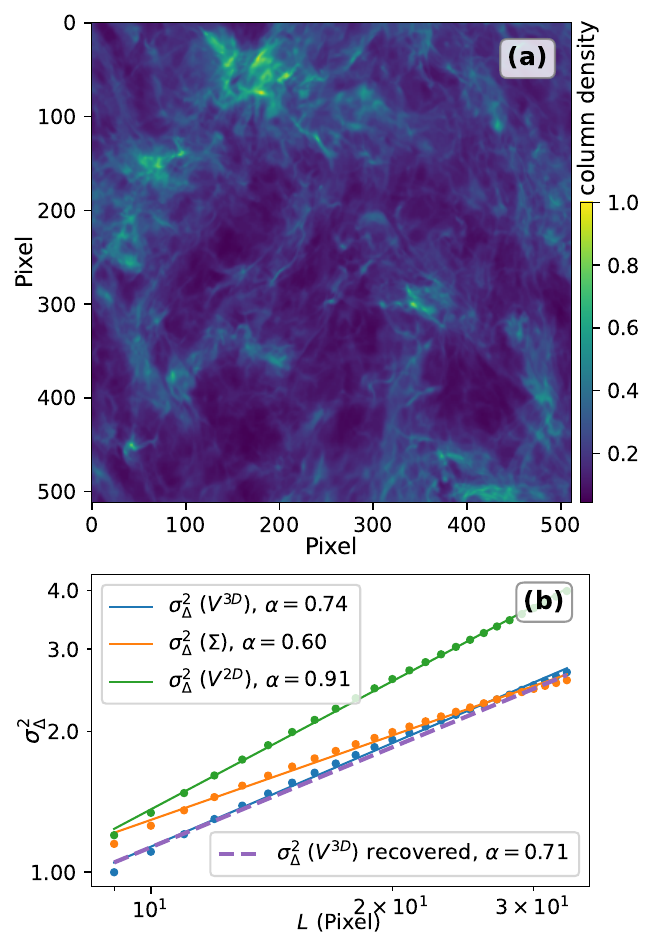}
    \caption{Same as Figure \ref{fig:simups_new} but for the new simulation (Sect. \ref{sec_arx_di}).}
    \label{fig:simups_new}
\end{figure}

\subsection{Archival Simulation of a Star-Forming System}\label{sec_arx_si}
To validate Eq.~\ref{eq_revi} under complex, star-forming conditions, we leverage the simulation dataset of filaments and star formation from \citet{2019MNRAS.485.4509L}. We select two control subregions (indicated by the white boxes in Figure~\ref{fig:simups}a) that remain unaffected by stellar feedback or active star formation. For each subregion, we calculate three distinct scaling parameters: the intrinsic value $\alpha_V^{\mathrm{3D}}$ (obtained via direct fitting of the 3D velocity data cube), $\alpha_V^{\mathrm{2D}}$ (derived from the density-weighted line-of-sight velocity map), and $\alpha_\Sigma^{\mathrm{2D}}$ (derived from the column density map). As expected, Eq.~\ref{eq_revi} holds robustly in both control regions, with the derived $\alpha_V^{\mathrm{3D}}$ deviating by less than 0.05 from the intrinsic value.

\subsection{Simulation of a Purely Turbulent System}\label{sec_arx_di}
To test Eq.~\ref{eq_revi} without stellar feedback biases, we run an idealized simulation ($512^3$ grid) using the \textsc{Orion2} code \citep{2012ApJ...745..139L}. The setup omits self-gravity to model a weakly magnetized, purely turbulent medium, initialized with a highly supersonic sonic Mach number of $\mathcal{M}_s = 9.4$ and an Alfvénic Mach number of $\mathcal{M}_A = 6.6$. For this isotropic system, the column density scaling parameter $\alpha_\Sigma$ remains less than unity. This confirms that spatial inhomogeneity must be explicitly accounted for when mapping 2D observational parameters to their intrinsic 3D counterparts. Crucially, Eq.~\ref{eq_revi} holds robustly for the synthetic velocity fields (Figure~\ref{fig:simups_new}), demonstrating that our methodology yields a reliable estimation of $\alpha_V^{\mathrm{3D}}$.

\end{appendix}
\end{document}